\documentclass[twocolumn,showpacs,preprintnumbers,amsmath,amssymb,floatfix]{revtex4}

\usepackage{graphicx}
\usepackage{bm}

\begin{document}

\preprint{APS/123-QED}

\title{Number theoretic example of scale-free topology inducing
    self-organized criticality}


\author{Bartolo Luque$^1$, Octavio Miramontes$^2$ and Lucas Lacasa$^{1,*}$}
\affiliation{$^1$Departamento de Matem\'atica Aplicada y
Estad\'istica, ETSI  Aeron\'auticos,  Universidad Polit\'ecnica de
Madrid,   Madrid  28040,   Spain\\\\$^2$Departamento  de   Sistemas
Complejos, Instituto de F\'isica, Universidad  Nacional Aut\'onoma
de M\'exico, 04510 DF, Mexico }

\date{\today}

\begin{abstract}
In this work we present a general mechanism by which simple dynamics
running on networks become self-organized critical for scale free
topologies. We illustrate this mechanism with a simple arithmetic
model of division between integers, the division model. This is the
simplest self-organized critical model advanced so far, and in this
sense it may help to elucidate the mechanism of self-organization to
criticality. Its simplicity allows analytical tractability,
characterizing several scaling relations. Furthermore, its
mathematical nature brings about interesting connections between
statistical physics and number theoretical concepts. We show how
this model can be understood as a self-organized stochastic process
embedded on a network, where the onset of criticality is induced by
the topology.
\end{abstract}

 \maketitle 


\noindent In the  late 80s Bak, Tang  and Wiesenfeld (BTW)
\cite{bak1987,bak1988} introduced the concept of Self-Organized
Criticality (SOC) as a mechanism explaining how multicomponent
systems can evolve naturally into barely stable self-organized
critical structures without external ``tuning" of parameters. This
single contribution sparkled an enormous theoretical and
experimental research interest in many areas of physics and
interdisciplinary science, and many natural phenomena were claimed
to exhibit SOC \cite{bak1996,jensen,SOC}. However, there was not a
general accepted definition of what SOC exactly is, and the
conditions under which it is expected to arise. In order to
disengage the mechanism of self-organization to criticality one
should likely focus on rather `simple' models, and in this sense
Flyvbjerg recently introduced the ``simplest SOC model" along with a
workable definition of the phenomenon \cite{Flyvbjerg1,Flyvbjerg2},
namely `a \emph{driven}, \emph{dissipative} system consisting of a
\emph{medium} through which \emph{disturbances} can propagate
causing a \emph{modification} of the medium, such that eventually,
the disturbances are \emph{critical}, and  the medium is {\it
modified no more} $-$in the statistical sense'.\\
\noindent On the other hand, in the last years it has been realized
that the dynamics of processes taking place on networks evidence a
strong dependence on the network's topology \cite{newmann,
barabasi}. Concretely, there exist a current interest on the
possible relations between SOC behavior and scale-free networks
\cite{barabasi}, characterized by power law degree distributions
$P(k)\sim k^{-\gamma}$, and how self-organized
critical states can emerge when coupling topology and dynamics \cite{interplay1, interplay2, interplay3, interplay4}.\\
\noindent In this work we introduce a rather simple and general
mechanism by which the onset of criticality in the dynamics of
self-organized systems is induced by the scale-free topology of the
underlying network of interactions. To illustrate this mechanism we
present a simple model, the division model from now on, based
uniquely in the division between integers. We show that this model
compliances with Flyvbjerg's definition of SOC and to our knowledge,
constitutes the simplest SOC model advanced so far that is also
analytically solvable. Interestingly, this model establishes
connections between statistical physics and number theory (see
\cite{web} for a complete bibliography on this topic).\\

\noindent In number theory, a primitive set of $N$ integers is the
one for which none of the set elements divide exactly any other
element \cite{erdos, primitive,abiertos1}. Consider an ordered set
of $M-1$ integers $\{2,3,4,..,M\}$ (notice that zero and one are
excluded, and that integers are not repeated), that we will name as
the pool from now on. Suppose that we have extracted $N$ elements
from the pool to form a primitive set. The division model proceeds
then by drawing integers at random from the remaining elements of
the pool and introducing them in the set. Suppose that at time $t$
the primitive set contains $N(t)$ elements. The algorithm updating
rules are the following:\\
\noindent (R1) Perturbation: an integer $a$ is drawn from the pool
at random and
introduced in the primitive set.\\
 \noindent (R2) Dissipation: if $a$ divides and/or is divided by say
$s$ elements of the primitive set, then we say that an instantaneous
division-avalanche of size $s$ takes place, and these latter
elements are returned to the pool, such that the set remains
primitive but with a new size $N(t+1)=N(t)+1-s$.\\
\noindent This process is then iterated, and we expect the primitive
set to vary in size and composition accordingly. The system is
\emph{driven} and \emph{dissipative} since integers are constantly
introduced and removed from it, its size
temporal evolution being characterized by $N(t)$.\\

\begin{figure}[htpb]
\centering \includegraphics[width=0.45\textwidth]{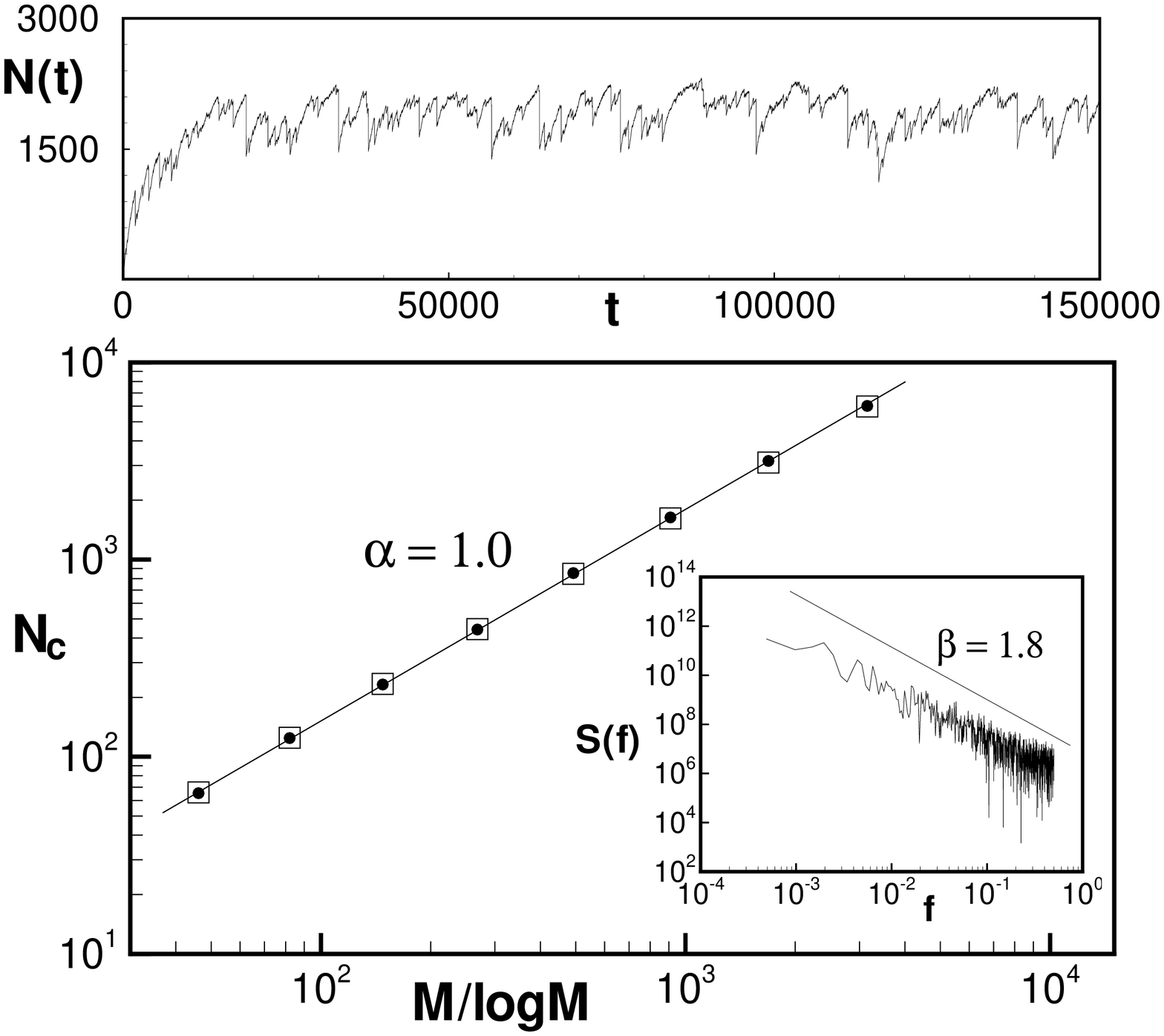}
\caption{\small{Upper figure: Single realization of the division
model showing the time evolution of the primitive set size $N(t)$
for a pool size $M=10^4$ and $N(0)=0$. Notice that after a
transient, $N(t)$ self-organizes around an average stable value
$N_c$, fluctuating around it. Bottom: (black dots) Scaling behavior
of the average stable value ${N_c}$ as a function
  of the system's characteristic size $M/\log M$. The best fitting provides  $ N_c\sim
 (M/\log M)^\gamma$, with $\gamma=1.05\pm0.01$. (squares) Scaling of $N_c$ as predicted by equation
 \ref{estacionario2}. Inner figure: plot in log-log of the power spectrum of $N(t)$, showing $f^{-\beta}$ noise
 with $\beta=1.80\pm0.01$ (this latter value is the average of $10^5$ realizations
 of $N(t)$ for $4096$ time steps after the transient and $M=10^4$).}}
\label{fig1}
\end{figure}

\begin{figure}[ht]
\includegraphics[width=0.45\textwidth]{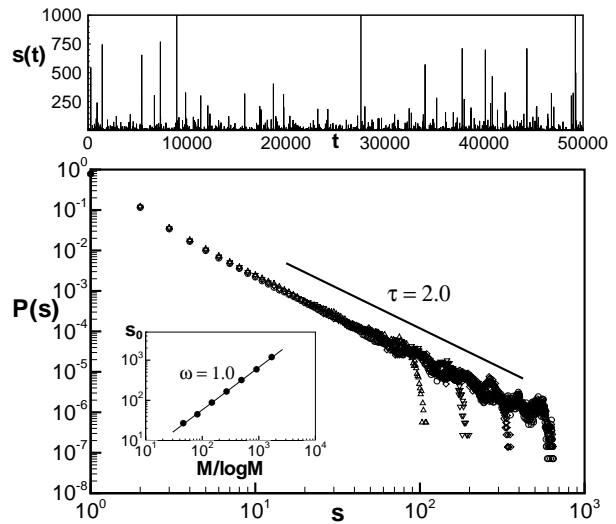}
\hspace{2cm} \caption{\small{Upper figure: Single realization of the
division model showing the time distribution of division-avalanches.
Bottom figure: Probability distribution $P(s)$ that a
division-avalanche of size $s$ takes place in the system, for
different pool sizes $M=2^{10}$ (triangles), $M=2^{11}$ (inverted
triangles), $M=2^{12}$ (diamonds) and $M=2^{13}$ (circles). In every
case we find $P(s)\sim s^{-\tau}\exp(s/s_0)$ with $\tau=2.0\pm0.1$.
Note that the power law relation evidences an exponential cut-off
due to finite size effects at particular values of $s_0$. Inner
figure: Scaling of the cut-off value $s_0$ as a function of the
system's characteristic size $M/ \log M$, with an exponent
$\omega=1.066\pm0.003$.}} \label{fig2}
\end{figure}

\noindent In order to unveil the dynamics undergoing in the model,
we have performed several Monte Carlo simulations for different
values of the pool size $M$. In the upper part of fig.\ref{fig1} we
have represented for illustration purposes a concrete realization of
$N(t)$ for $M=10^4$ and $N(0)=0$. Note that after a transient,
$N(t)$ self-organizes around an average stable value $N_c$,
fluctuating around it. In the inner part of the bottom fig.
\ref{fig1}, we have plotted in log-log the power spectrum of $N(t)$:
the system evidences $f^{-\beta}$ noise, with $\beta=1.80\pm0.01$.
The former fluctuations are indeed related to the fact that at each
time step a new integer extracted from the pool enters the primitive
set (external driving R1). Eventually (according to rule R2), a
division-avalanche can propagate and cause a {\it modification} in
the size and composition of the primitive set. These avalanches
constitute the {\it disturbances} of the system. In fig.\ref{fig2}
(up) we have represented an example of the avalanche's size
evolution in time. In the same figure (bottom) we show the
probability $P(s)$ that a division-avalanche of size $s$ takes
place, for different pool sizes $M$. These latter distributions are
power laws $P(s)\sim s^{-\tau}\exp(s/s_0)$ with $\tau=2.0\pm0.1$:
disturbances are thus critical. Observe that the power law relation
suffers from a crossover to exponential decay at a cut-off value
$s_0$ due to finite size effects (pool is finite), and that the
location of these cut-offs scales with the system's characteristic
size $s_0\sim(M/\log M)^\omega$ with $\omega=1.066\pm0.003$, what is
typically characteristic of a finite size critical state
\cite{jensen} (this characteristic size will be explained later in
the text). We can conclude that according to Flyvbjerg's definition
\cite{Flyvbjerg1}, the division model exhibits SOC.
Division-avalanches lead the system to different marginally stable
states, that are nothing but primitive sets of different sizes and
composition. Accordingly, for a given pool $[2,M]$, these time
fluctuations generate a
stochastic search in the configuration space of primitive sets.\\

\begin{figure}[h]
\includegraphics[width=0.5\textwidth]{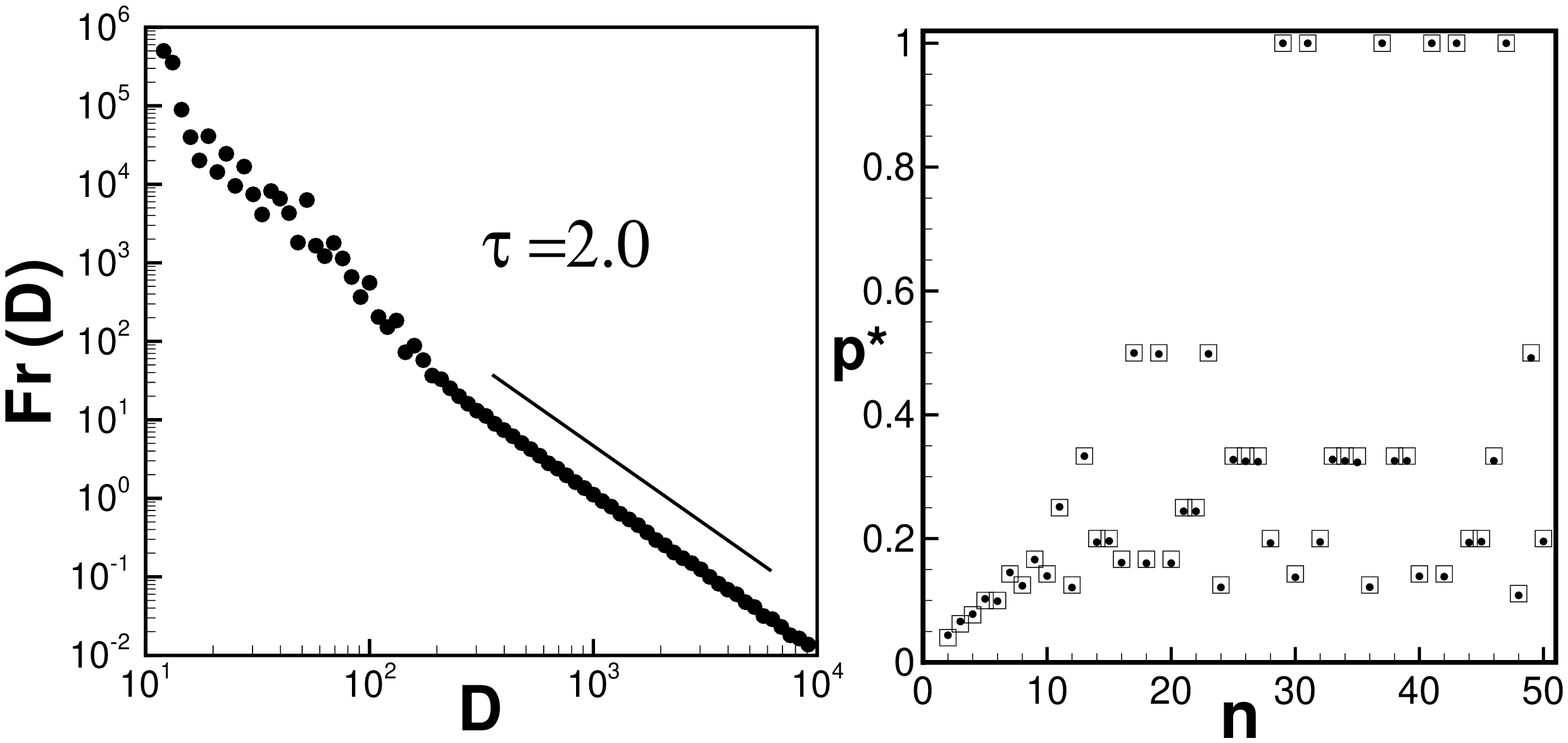}
\hspace{2cm} \caption{Left: Histogram of the amount of integers in
$[2,10^6]$ that have $D$ divisors. The histogram have been smoothed
(binned) to reduce scatter. The best fitting provides a power law
$P(D)\sim D^{-\tau}$ with $\tau=2.01\pm0.01$, in agreement with
$P(s)$ (see the text). Right: (black dots) Stationary survival
probability of integer $n$ in a primitive set for a pool size
$M=50$, obtained from Monte Carlo simulations of the model over
$10^6$ time steps (a preliminary transient of $10^4$ time steps was
discarded). (squares) Theoretical prediction of these survival
probabilities according to equation \ref{estacionario}.}
\label{fig3}
\end{figure}

\noindent In what follows we discuss analytical insights of the
problem. Consider the {\it divisor function}
\cite{libro_teoria_numeros} that provides the number of divisors of
$n$, excluding integers $1$ and $n$:
\begin{equation}
d(n)=\sum_{k=2}^{n-1}{\bigg(
  \bigg\lfloor\frac{n}{k}\bigg\rfloor-\bigg\lfloor\frac{n-1}{k}\bigg\rfloor\bigg)},
\end{equation}
where $\lfloor\ \rfloor$ stands for the integer part function. The
average number of divisors of a given integer in the pool $[2,M]$ is
then:
\begin{eqnarray}
&&\frac{1}{M-1}\sum_{n=3}^M{d(n)}=\frac{1}{M-1}\sum_{k=2}^M{\bigg\lfloor\frac{M}{k}\bigg\rfloor}
\simeq \sum_{k=2}^M{\frac{1}{k}}\nonumber\\
&&\simeq\log{M}+2(\gamma-1)+\textit{O}\bigg(\frac{1}{\sqrt{M}}\bigg).
\end{eqnarray}
Accordingly, the mean probability  that two numbers $a$ and $b$
taken at random from $[2,M]$ are divisible is approximately
$P=Pr(a|b)+Pr(b|a)\simeq 2\log{M}/M$. Moreover, if we assume that
the $N$ elements of the primitive set are uncorrelated, the
probability that a new integer generates a division-avalanche of
size $s$ is on average $(2\log M/M)N$. We can consequently build a
mean field equation for the system's evolution, describing that at
each time step an integer is introduced in the primitive set and a
division-avalanche of mean size $(2\log M/M)N$ takes place:
\begin{equation}
N(t+1)=N(t)+1-\bigg(\frac{2\log M}{M}\bigg)N(t), \label{master}
\end{equation}
\noindent whose fixed point $N_c=M/(2\log{M})$, the stable value
around which the system self-organizes, scales with the system's
size as
\begin{equation}
N_c(M)\sim \frac{M}{\log{M}}. \label{scaling}
\end{equation}
Hitherto, we can conclude that the system's characteristic size is
not $M$ (pool size) as one should expect in the first place, but
$M/\log M$.
This scaling behavior has already been
noticed in other number-theoretic models evidencing collective
phenomena \cite{luque1,lacasa}.
In fig.\ref{fig1} we have plotted (black dots) the values of $N_c$
as a function of the characteristic size $M/\log M$ provided by
Monte Carlo simulations of the model for different pool sizes
$M=2^8, 2^9,...,2^{15}$ ($N_c$ has been estimated averaging $N(t)$
in the steady state). Note that the scaling relation \ref{scaling}
holds, however the exact numerical values $N_c(M)$ are
underestimated by eq.\ref{master}. This is reasonable since we have
assumed that the primitive set elements are uncorrelated, what is
obviously not the case: observe for instance that any prime number
$p\geqslant \lfloor M/2\rfloor$ introduced in the primitive set will
remain there forever. Fortunately this drawback of our mean field
approximation can be improved by considering the function $D(n)$
that defines the exact number of divisors that a given integer $n
\in [2,M]$ has, i.e. the amount of numbers in the pool that divide
or are divided by $n$:
\begin{equation}
D(n)=d(n)+\bigg\lfloor\frac{M}{n}\bigg\rfloor-1.
\end{equation}
Define $p_n(t)$ as the probability that the integer $n$ belongs at
time $t$ to the primitive set. Then, we have
\begin{equation}
p_n(t+1)=\bigg(1-\frac{D(n)}{M-N(t)}\bigg) p_n(t)+
\frac{1}{M-N(t)}\big(1-p_n(t)\big),
\end{equation}
that leads to a stationary survival probability in the primitive
set:
\begin{equation}
p_n^*=\frac{1}{1+D(n)}. \label{estacionario}
\end{equation}
In Fig.\ref{fig3} (right) we depict the stationary survival
probability of integer $n$ (black dots) obtained through numerical
simulations for a system with $M=50$, while squares represent the
values of $p_n^*$ as obtained from the eq.\ref{estacionario}. Note
that there exists a remarkable agreement. We now can proceed to
estimate the critical size values $N_c(M)$ as:
\begin{equation}
N_c(M) \approx \sum_{n=2}^Mp_n^*= \sum_{n=2}^M{\frac{1}{1+D(n)}}.
\label{estacionario2}
\end{equation}
In fig.\ref{fig1} we have represented (squares) the values of
$N_c(M)$ predicted by eq.\ref{estacionario2}, showing good
agreement with the numerics (black dots).\\

\noindent Finally, previous calculations point out that system's
fluctuations, i.e. division-avalanches distribution $P(s)$ is
proportional to the percentage of integers having $s$ divisors. In
order to test this conjecture, in fig.\ref{fig3} (left) we have
plotted a histogram describing the amount of integers having a given
number of divisors, obtained from computation of $D(n)$ for
$M=10^6$. The tail of this histogram follows a power law with
exponent $\tau=2.0$. This can be proved analytically as it follows:
the responsible for the tail of the preceding histogram are those
numbers that divide many others, i.e. rather small ones ($n\ll M$).
A small number $n$ divides typically
$D(n)\simeq\lfloor\frac{M}{n}\rfloor$. Now, how many `small numbers'
have $D(n)$ divisors? The answer is $n$, $n+1$,..., $n+z$ where
\begin{equation}
\bigg\lfloor\frac{M}{n}\bigg\rfloor=
\bigg\lfloor\frac{M}{n-1}\bigg\rfloor=...=\bigg\lfloor\frac{M}{n-z}\bigg\rfloor.
\end{equation}
The maximum value of $z$ fulfills $\frac{M}{n-z}-\frac{M}{n}=1$,
that is $z\simeq n^2/M$. The frequency of
 $D(n)$ is thus $fr(D(n))=n^2/M$, but since $s\equiv D(n)\simeq
 M/n$, we get $fr(s)\sim Ms^{-2}$, and finally normalizing, $P(s)\sim
 s^{-2}$.\\

\noindent Coming back to the Flyvbjerg's definition of SOC, which is
the \emph{medium} in the division model? Observe that the process
can be understood as embedded in a network, where nodes are
integers, and two nodes are linked if they are exactly divisible.
The primitive set hence constitutes a subset of this network, that
is dynamically modified according to the algorithm's rules. The
degree of node $n$ is $D(n)$, and consequently the degree
distribution $P(k)\sim k^{-2}$ is scale-free. Hence the SOC
behavior, which arises due to the divisibility properties of
integers, can be understood as a sort of anti-percolation process
taking place in this scale-free network. Observe that the division
model is a particular case of a more general class of self-organized
models: a network with $M$ nodes having two possible states
(\emph{on/off}) where the following dynamics runs: (R1)
perturbation: at each time step a node in the state \emph{off} is
randomly chosen and switched \emph{on}, (R2) dissipation: the $s$
neighbors of the perturbed node that were in the state \emph{on} in
that time step are switched \emph{off}, and we say that an
instantaneous avalanche of size $s$ has taken place. $N(t)$ measures
the number of nodes in the state \emph{on} as a function of time.
Its evolution follows a mean field equation that generalizes eq.
\ref{master}:
\begin{equation}
N(t+1)=N(t)+1-\frac{\langle k\rangle}{M}N(t),
\end{equation}
where $\langle k\rangle$ is the network's mean degree. Accordingly,
in every case $N(t)$ will self-organize around an average value
$N_c(M)$. Within regular or random networks, fluctuations
(avalanches) around $N_c(M)$ will follow a Binomial or Poisson
distribution respectively. However, when the network is scale free
with degree distribution $P(k)\sim k^{-\gamma}$, fluctuations will
follow a power law distribution $P(s)\sim s^{-\tau}$ with
$\tau=\gamma$, and the dynamics will consequently be SOC. In this
sense, we claim
that scale-free topology induces criticality.\\
\noindent Some questions concerning this new mechanism can be
depicted, namely: which is the relation between the specific
topology of scale-free networks and the power spectra of the
system's dynamics? Which physical or natural systems evidence this behavior?\\
\noindent With regard to the division model, the bridge between
statistical physics and number theory should also be investigated in
depth. This includes possible generalizations of this model to other
related sets such as $k$-primitive sets \cite{future}, where every
number divides or is divided by at least $k$ others ($k$ acting as a
threshold parameter), to relatively primitive sets \cite{coprime}
and to cross-primitive sets \cite{primitive} (where this will
introduce coupled SOC models). From the computational viewpoint
\cite{SFI}, properties of the model as a primitive set generator
should also be studied. Of special interest is the task of
determining the maximal size of a $k$-primitive set \cite{primitive,
future}, something that can be studied within the division model
through extreme value theory \cite{bak1996}.\\
\small{The authors appreciate valuable comments from P. Miramontes
and D. Boyer, financial support by UNAM-DGAPA grant IN-118306, grant
FIS2006-08607 from the Spanish ministry of Science, the ``Tomas
Brody Lectureship 2007" awarded to BL and the Instituto de Fisica at
UNAM for support and hospitality. OM also thanks the hospitality of
the DMAE of the ETSI-Aeronauticos at UPM, during a 2008 academic
visit.}

$^*$\emph{Electronic address}: lucas@dmae.upm.es

\end{document}